\documentclass[11pt]{article}
\usepackage{graphics}
 
\begin{document}

\def\oti{{\otimes}}
\def\bra#1{{\langle #1 |  }}
\def\lb{ \left[ }
\def\rb{ \right]  }
\def\tilde{\widetilde}
\def\bar{\overline}
\def\hat{\widehat}
\def\*{\star}

\def\({\left(}		\def\BL{\Bigr(}
\def\){\right)}		\def\BR{\Bigr)}
	\def\BBL{\lb}
	\def\BBR{\rb}
%
%
\def\bb{{\bar{b} }}
\def\ab{{\bar{a} }}
\def\zb{{\bar{z} }}
\def\zbar{{\bar{z} }}
\def\frac#1#2{{#1 \over #2}}
\def\inv#1{{1 \over #1}}
\def\half{{1 \over 2}}
\def\d{\partial}
\def\der#1{{\partial \over \partial #1}}
\def\dd#1#2{{\partial #1 \over \partial #2}}
\def\vev#1{\langle #1 \rangle}
\def\ket#1{ | #1 \rangle}
\def\rvac{\hbox{$\vert 0\rangle$}}
\def\lvac{\hbox{$\langle 0 \vert $}}
\def\2pi{\hbox{$2\pi i$}}
\def\e#1{{\rm e}^{^{\textstyle #1}}}
\def\grad#1{\,\nabla\!_{{#1}}\,}
\def\dsl{\raise.15ex\hbox{/}\kern-.57em\partial}
\def\Dsl{\,\raise.15ex\hbox{/}\mkern-.13.5mu D}
%
%
\def\th{\theta}		\def\Th{\Theta}
\def\ga{\gamma}		\def\Ga{\Gamma}
\def\be{\beta}
\def\al{\alpha}
\def\ep{\epsilon}
\def\vep{\varepsilon}
\def\la{\lambda}	\def\La{\Lambda}
\def\de{\delta}		\def\De{\Delta}
\def\om{\omega}		\def\Om{\Omega}
\def\sig{\sigma}	\def\Sig{\Sigma}
\def\vphi{\varphi}
%
%
\def\CA{{\cal A}}	\def\CB{{\cal B}}	\def\CC{{\cal C}}
\def\CD{{\cal D}}	\def\CE{{\cal E}}	\def\CF{{\cal F}}
\def\CG{{\cal G}}	\def\CH{{\cal H}}	\def\CI{{\cal J}}
\def\CJ{{\cal J}}	\def\CK{{\cal K}}	\def\CL{{\cal L}}

\def\CM{{\cal M}}	\def\CN{{\cal N}}	\def\CO{{\cal O}}
\def\CP{{\cal P}}	\def\CQ{{\cal Q}}	\def\CR{{\cal R}}
\def\CS{{\cal S}}	\def\CT{{\cal T}}	\def\CU{{\cal U}}
\def\CV{{\cal V}}	\def\CW{{\cal W}}	\def\CX{{\cal X}}
\def\CY{{\cal Y}}	\def\CZ{{\cal Z}}

\def\rvac{\hbox{$\vert 0\rangle$}}
\def\lvac{\hbox{$\langle 0 \vert $}}
\def\comm#1#2{ \BBL\ #1\ ,\ #2 \BBR }
\def\2pi{\hbox{$2\pi i$}}
\def\e#1{{\rm e}^{^{\textstyle #1}}}
\def\grad#1{\,\nabla\!_{{#1}}\,}
\def\dsl{\raise.15ex\hbox{/}\kern-.57em\partial}
\def\Dsl{\,\raise.15ex\hbox{/}\mkern-.13.5mu D}
\def\beq{\begin {equation}}
\def\eeq{\end {equation}}

\title{Quantum Rate-Distortion Theory for I.I.D. Sources}
\author { Igor Devetak\footnote{Electronic address: igor@ece.cornell.edu}
  $~$and Toby Berger\\ 
  $~$\\
  Department of Electrical and Computer Engineering\\
  $~$\\
  Cornell University, Ithaca, New York 14853}
   
\maketitle

\begin{abstract}
\noindent
We formulate quantum rate-distortion theory in the most general
setting where classical side information is included in the tradeoff.
Using a natural distortion measure based on entanglement fidelity
and specializing to the case of an unrestricted classical side channel,
we find the exact quantum rate-distortion function for a source of isotropic
qubits. An upper bound we believe to be exact
is found in the case of biased sources.
We establish that in this scenario optimal rate-distortion codes produce no 
entropy exchange with the environment of any individual qubit.
\end{abstract}
  

\section{Introduction}
 
The quantum lossless source coding theorem specifies
the minimum rate, called the {\it entropy} and
measured in code qubits per source qubit, to which 
a quantum source can be compressed subject to the requirement that
the source qubits can be recovered \emph{perfectly} from the code qubits.
In realistic applications we may be able to 
tolerate imperfect recovery of the source
qubits at the receiver, in which 
case we would seek
to minimize the rate required to achieve a
specified level of distortion.  Equivalently, 
we may be required to use a rate $R$ less than the entropy of the source,
in which case we would seek
to minimize the distortion subject to this rate constraint.
Here, the distortion measure is 
a user-defined function of the input and the reconstruction
the precise form
of which depends on the nature of the application.  
\vskip 0.1in\noindent
Analysis of the tradeoff
between rate and distortion is the subject matter 
of \emph{rate-distortion theory}.
Classical rate-distortion theory \cite{berger} is an important and 
fertile area in information theory.  Considering that coding theorems for both
noiseless \cite{nono} and noisy \cite{noisy} quantum {\it channels} have been 
established some years ago, it is surprising that little effort has
been put into developing quantum rate-distortion theory.
The purpose of this paper is to 
fill that gap.
\vskip 0.1in\noindent
To be completely general one must allow for a classical side channel
containing information gathered while manipulating the source qubits,
and include the corresponding classical rate $r$, 
measured in bits per source qubit, in the tradeoff.
It has been shown in \cite{bj} that at zero distortion no classical
side information can help reduce the quantum rate below the von Neumann
entropy of the source. This turns out not to be the case 
for positive distortion $d$. Therefore one must speak of a $2$-dimensional
tradeoff manifold $R(d,r)$. 
Here we introduce this general formulation for the first time.
However, we focus mainly on the scenario
of unrestricted classical side information, i.e. $r = \infty$, 
and refer to $R(d) \equiv R(d,\infty)$ as the rate-distortion
function. This clearly provides a lower bound on achievable $R$ for the same 
distortion $d$ but restricted classical rate $r$. 

In classical information theory the rate-distortion function 
has the simple form
\beq
   R(D) = \min_{Y: E_{X,Y} d(X,Y) \leq D} I(X;Y), 
\label{eq:e0}
\eeq 
where $X$ is a random variable distributed like a typical source letter,
$Y$ is a random variable jointly distributed with $X$ that is
used to construct approximations to the source output and  ranges over 
an alphabet possibly different from the source alphabet,
$E_{X,Y}$ denotes expectation with respect
to the joint distribution of $X$ and $Y$,
$d(\cdot,\cdot)$ is a suitably defined distortion function,
and $I(X;Y)$ is the average mutual information between $X$ and $Y$.
The relevant 
information-like quantity playing the role in the quantum channel capacity
formula is the coherent information $I_c(\rho,\CE) $ \cite{sn} to be defined
in the next section.
The natural first impulse is to assume that the
same quantity should appear in quantum rate-distortion theory.
Indeed, Barnum \cite{barnum}
has derived a lower bound on $R(d,0)$ based on coherent information.
This bound is far from tight, however, since the coherent 
information often is negative for distortions considerably smaller that that
which can be achieved with the receiver is sent no qubits at all. 
(A comparable problem does not occur in channel capacity calculations because
the maximization procedure invoked there ensures 
positivity.)  In view of this we pursue quantum 
rate-distortion from first principles using a natural distortion
measure based on entanglement fidelity that was introduced in \cite{barnum}.

We define the problem in Section 2,
wherein we also provide
relevant background on quantum operations, entropies and fidelity measures.
In Section 3 we find the rate-distortion function for a restricted class
of coding procedures; in Section 4 we argue that the optimum coding scheme
belongs to this class.  Section 5 describes
a simple physical realization of the optimal coding procedure.
Speculations are left for the final section.
\section{Definitions}
Let us recall some basic definitions of quantum information theory
\cite{sch}, \cite{bns}.
A general quantum information source is described by a 
density matrix $\rho^Q$ of a quantum
system $Q$. This density matrix may result from  the system being prepared 
in certain pure states with respective probabilities. Alternatively, we
may view our quantum system $Q$ as a part of a larger system $RQ$
which includes a \emph{reference system} $R$ which always may be constructed
such that the overall
state is pure $\ket{\Psi^{RQ}}$ and $\rho^Q$ results from restricting
to $Q$, i.e.,
\beq
\rho^Q = tr_R(\ket{\Psi^{RQ}} \bra{\Psi^{RQ}})~.
\eeq
\vskip 0.1in\noindent
Next consider a quantum process acting on the source $\rho^Q$
\beq
\rho^Q \rightarrow  \hat{\CE}(\rho^Q) \equiv 
\frac{\CE(\rho^Q)}{tr(\CE(\rho^Q))},  \label{eq:e1}
\eeq
with a general quantum operation ${\CE}$ of the form 
\beq
\CE(\rho^Q) = \sum_{i=1}^k A_i \rho^Q A_i^\dagger~.
 \label{eq:e2} 
\eeq 
Note that the action of $\CE$ is completely determined by the set  
of operation elements $\{ A_i \}$. 
A useful way to think about the quantum process is to embed
$RQ$ into an even larger space $RQE$ by adding an \emph{environment}
$E$, initially in a pure state $\ket{s}$ and hence decoupled from 
$RQ$. Then a well-known representation theorem \cite{sch}, \cite{bns} 
states that a general quantum process $\CE$ may be realized by performing
a unitary transformation $U^{QE}$ entangling $Q$ and $E$, followed
by projecting via $P^E$ onto the environment alone, and then tracing out 
$R$ and $E$; i.e.,
\beq
\CE(\rho^Q) = c ~tr_{RE}(P^E U^{QE}(\ket{\Psi^{RQ}} \bra{\Psi^{RQ}}
\otimes \ket{s} \bra{s}) U^{QE \dagger} P^E ),
\eeq
where $c$ is a positive constant. Although the theorem refers to
a mathematical construction, it provides physical insight.
For instance, it enables one to define the entropy exchange
\cite{sch}, \cite{noisy}
\beq
S_e(\rho^Q,\CE) \equiv S(\rho^{E'}) = S(\rho^{RQ'}) 
\label{eq:ex}
\eeq
Here $S(\sigma) \equiv - tr (\sigma \log_2 \sigma)$ is the von Neumann
entropy and  
$\rho^{E'} $ and $\rho^{RQ'}$ denote the states of $E$ and $RQ$, respectively,
after the operation. The equality in (\ref{eq:ex}) holds because the whole
system $RQE$ remains in a pure state after the process, as a consequence of
which $S_e(\rho^Q,\CE)$ measures
the amount of ``disorder'', or ``noise'',
introduced into the system $RQ$ by virtue of its having
become entangled with $E$, and vice versa. 
\vskip 0.1in\noindent
A convenient expression
in terms of the original operation elements $\{A_i\}$ is
\beq
S_e(\rho^Q,\CE)  =  S(W) = - tr(W \log_2 W)
\eeq  
\vskip 0.1in\noindent
with 
\beq
W_{ij} = \frac{tr(A_i \rho^Q A_j^\dagger)}{tr(\CE(\rho^Q)) }
\eeq
\vskip 0.1in\noindent
Observe that if there is only one operation element (or, equivalently,
if they are all the same), then the entropy exchange is zero.
The noise interpretation of $S_e$ is also evident from the formula
for coherent information,
\beq
I_c(\rho^Q,\CE) = S(\hat{\CE}(\rho^Q)) -  S_e(\rho^Q,\CE),
\eeq
that appears in the channel capacity formula. Comparing $I_c(\rho^Q,\CE)$
to its classical counterpart $I(X;Y) = H(Y) - H(Y|X)$, we see that
$S_e(\rho^Q,\CE)$ plays a role analogous to the noise term,
$H(Y|X)$.
\vskip 0.1in\noindent
We end this brief review with the definition of \emph{
entanglement fidelity}, denoted by $F_e(\rho^Q,\CE)$ and defined by
\beq
F_e(\rho^Q,\CE) =  \bra{\Psi^{RQ}} (I_R \otimes \CE)
(\ket{\Psi^{RQ}}\bra{\Psi^{RQ}} )  \ket{\Psi^{RQ}}.
\eeq 
The entanglement fidelity tells us how well the system's state
and the system's entanglement with its surroundings
$R$, which do not participate directly in the quantum process, 
are preserved under the operation in question.
Like any meaningful quantity it has an expression which is manifestly
independent of which purification $R$ is employed, namely
\beq
F_e(\rho^Q,\CE) = \frac{\sum_i | tr(A_i \rho^Q) | ^2 }{tr(\CE(\rho^Q)) }
\eeq 
\vskip 0.1in\noindent
We now augment Barnum's formulation of the $r = 0$ case 
\cite{barnum} to allow for classical side information.
First we restrict attention to i.i.d. 
sources with density matrix $\rho$,
so that  $\rho^{(n)} \equiv \rho^{\otimes n}$. An $(n,R,r)$ \emph{
rate-distortion code}
consists of an encoding operation $\CC^{(n)}$ from the source space 
$\rho^{(n)}$ to a block of 
$ \lfloor  nR \rfloor$ qubits and $ \lfloor  nr \rfloor$ bits
(henceforth abbreviated to $nR$ and $nr$ respectively),
and a decoding operation 
$\CD^{(n)}$ acting in the reverse direction. Here $R \leq 1$, so in effect
we are compressing the $n$ source qubits to $nR$ qubits and then decompressing
them back to $n$ qubits with the help of $nr$ bits of information 
gathered during the compression phase, in an attempt to recover the original with
the maximum possible fidelity consistent with the values of $R$ and $r$. 
\vskip 0.1in\noindent
For the rate-distortion code $( \CC^{(n)}, \CD^{(n)} )$ 
Barnum defines a natural distortion 
based on entanglement fidelity, namely 
\beq
d_e(\rho^{(n)},  \CD^{(n)} \circ \CC^{(n)} ) \equiv \sum_{\alpha=1}^n \inv{n} 
(1 -  F_e (\rho, \CT^\alpha ))
\eeq
with $\CT^\alpha$ being the marginal operation on the $\alpha$-th copy of
$\rho$ induced by the encoding-decoding operation,
\beq
\CT^\alpha (\sigma) \equiv tr_{1,...,\alpha-1,\alpha+1,...,n} \CD^{(n)}
\circ \CC^{(n)} 
(\rho \otimes \rho \cdots \otimes \rho \otimes  \sigma \otimes \rho 
\cdots  \otimes \rho  ).
\eeq
\vskip 0.1in\noindent
We say that a rate-distortion triplet $(R,r,d)$ is \emph{achievable} 
for a given $\rho$
iff there exists a sequence of $(n,R,r)$ rate-distortion codes 
$( \CC^{(n)}, \CD^{(n)} )$ such that
\beq
\lim_{n \rightarrow \infty} d_e(\rho^{(n)} ,\CD^{(n)} \circ \CC^{(n)}) 
\leq d
\eeq 
Then the \emph{rate-distortion manifold} $R(d,r)$ is defined as the infimum of
all $R$ for which $(R,r,d)$ is achievable.  
\vskip 0.1in\noindent

 In the following we approach the problem of finding $R(d,r)$ from first 
principles.
With no loss of generality the encoding procedure may be divided into
two steps.
In the first step the encoder manipulates blocks of qubits of size $n$ via some
quantum operation
$ \CE(\rho^{(n)}) = \sum_{i=1}^k A_i \rho^{(n)} A_i^\dagger$.
For $\CE$ to be physical its operation elements $\{ A_i \}$ must satisfy the 
trace preserving condition $ \sum_{i = 1}^k A_i^\dagger A_i = I$.
Define quantum operations $\CE_{A_i}(\rho^{(n)}) = 
A_i \rho^{(n)} A_i^\dagger$.
A given decomposition \{$A_i$\} of unity implies that the probability
that the non-trace preserving operation ${\CE}_{A_i}$ is the one that will be performed is
$\lambda_i = tr(\CE_{A_i}(\rho^{(n)}))$.
Quantum mechanics forbids the encoder to have {\it control} over which of
the $k$ operations will get performed, but afterwards the encoder can
obtain {\it information} about which one actually took place. 
This information is embodied in the index random variable
$I$ taking integer values $i$, $1 \leq i \leq k$ with respective probabilities $\lambda_i$.
In general some or all of this information may be available to the decoder, 
embodied in the random variable $J = f(I)$, a deterministic function of $I$. 
Further define 

\beq
\bar{S} = E_J S(E_{I|J} \hat{\CE}_{A_I}(\rho^{(n)})) \,\, ,
\eeq
 
the average output von Neumann entropy conditional on the knowledge of $J$
(i.e. from the point of view of somebody who knows the value of $J$ but not
the value of $I$).
Given $R$ and $r$, the goal is to choose $\CE$ and $f$ so that 
the distortion is minimized while keeping $\bar{S} \leq nR$ and $H(J) \leq nr$.

In the second step we take a large number $N$ of such blocks, 
group them according
to the value of $J$, and process each group
in the standard lossless coding way \cite{nono,chuang}
in order to get a string of at most $NnR$
qubits in the limit of large $N$. 
The decoding procedure is just reversing
the second step, which the lossless coding theorem assures us can
be done with effectively perfect fidelity in the limit as $N \to \infty$
(for fixed $n$),
and using the $Nnr$ bits of classical information about the values of
$J$ for each block so the decoder may unscramble them properly.
Finally, the rate-distortion function will be achieved in the limit of large $n$,
as well as large $N$.

Since the distortion depends only on the operation elements
${A_i}$, the choice of $f$ only affects the tradeoff between $R$ and $r$.
Using the concavity of von Neumann entropy \cite{wehrl} and the fact
that $E_J E_{I|J} = E_I$, we have the following inequalities:

\beq
  E_I S( \hat{\CE}_{A_I}(\rho^{(n)})) \leq
  \bar{S} \leq S( E_I \hat{\CE}_{A_I}(\rho^{(n)})) =
  S(\CE(\rho^{(n)}))    
\eeq  

The upper bound is attained when $f = const$, i.e. when no classical side
information is allowed. 
The lower bound on $\bar{S}$ is attained when $f$ is the identity map,
in which case $H(J) = H(I) = - \sum_{i=1}^k \la_i \log_2 \la_i$ is maximum.
An intuitive argument for the latter is that, 
from the point of view of the decoder, 
only single element operations ${\CE}_{A_i}$ have been performed; 
these in turn have zero entropy exchange, which we interpreted as noise.
Whenever the decoder lacks information about the value of $I$, the entropy
exchange of the block is strictly positive.

We shall henceforth concentrate on the case of maximal classical 
rate $r$, thus reducing the problem to finding the tradeoff 
function $R_n(d)$ between $\bar{S} = \sum_{i = 1}^k \lambda_i 
\space S \( \hat{\CE}_{A_i}(\rho^{(n)}) \)$ and the distortion 
$d_e(\rho^{(n)},  \CE )$. The rate distortion function 
is given by the limit $R(d) = \lim_{n \rightarrow \infty} R_n(d)$. 
In the next section we analyze the $n = 1$ case.
Subsequently we demonstrate the perhaps surprising fact 
that $n = 1$ already attains the $R(d)$ curve.

\section{The rate-distortion function for $n = 1$ }
\newtheorem{thm1}{Theorem}
\newtheorem{thm2} [thm1]{Theorem}
\newtheorem{thm3}[thm1]{Theorem}
\newtheorem{thm4}[thm1]{Theorem}
\newtheorem{lemma1} {Lemma}
\newtheorem{lemma'}[lemma1]{Lemma}
\newtheorem{lemma'2}[lemma1]{Lemma}  
Let us temporarily restrict attention to $k=1$,  so that
(\ref{eq:e2}) becomes $\CE(\sigma) =  A \sigma A^\dagger$, 
and also temporarily ignore the trace-preserving constraint.
First a technical lemma:
\begin{lemma1}
Let $\Delta$ and $\Lambda$ be positive diagonal matrices whose 
diagonal elements are given in a non-ascending order. Then for any unitary 
$U$ and $V$ the
inequality $ |tr( U \Delta V \Lambda) | \leq tr(\Delta \Lambda)$ holds.
\end{lemma1}
$\mathbf{Proof }  $ \space \space 
Consider the Cauchy-Schwartz inequality for the Hilbert-Schmidt
inner product \cite{bns} $\langle A , B \rangle \equiv tr(A B^\dagger)$, 
namely
\beq
|tr(A B^\dagger )|^2 \leq tr(A A^\dagger ) tr(B B^\dagger ).
\label{eq:CS}
\eeq
Since $\Delta$ and $\Lambda$ are positive we have 
$\Delta = \sqrt{\Delta \Delta^\dagger}$ and $\Lambda =
\sqrt{\Lambda \Lambda^\dagger}$.
Setting $A = \sqrt{\Delta} V  \sqrt{\Lambda}$ and 
$B = \sqrt{\Delta} U^\dagger \sqrt{\Lambda}$, 
we find that
\beq
|tr( U \Delta V \Lambda) |^2 \leq  tr( U \Delta U^\dagger \Lambda) 
tr( V \Delta V^\dagger \Lambda),
\eeq
so the lemma will hold for general unitary $U$ and $V$ provided it holds
when $V = U^\dagger$. 
Next, denote the  elements of $U$ and diagonal elements 
of $\Delta$ and $\Lambda$ 
by $\{u_{ij}\}$, $\{\delta_i\}$ and $\{\lambda_i\}$, respectively. 
Defining the matrix $P$ with elements $p_{ij} =  |u_{ij}|^2$, we have 
\beq
tr( U \Delta U^\dagger \Lambda)  =  
\sum_{i,j} u_{ij} \delta_j u_{ij}^* \lambda_i 
= \sum_{i,j} p_{ij} \delta_j \lambda_i  . 
\eeq
Since elements of each row and column of $P$ add up to $1$, $P$ is a stochastic
matrix, and hence a convex combination of permutation matrices \cite{wehrl}. So
the maximum value of $tr( U \Delta U^\dagger \Lambda)$ is equal to
$\sum_i \delta'_i \lambda_i$ with $\delta'_i$ a permutation of the $\delta_i$.
By Chebyshev's inequality $P = I$ corresponds
to the optimum permutation; this is especially easy to see for
$2 \times 2$ matrices for which the ordering condition implies 
$(\lambda_1  - \lambda_2) (\delta_1 - \delta_2) \geq 0$, or
$ \lambda_1 \delta_1 + \lambda_2 \delta_2  \geq 
\lambda_1 \delta_2 +  \lambda_2 \delta_1  $.
Therefore $U = V = I$ maximizes  $|tr( U \Delta V \Lambda) |$; the
Lemma is proved.
\begin{thm1}
For all single qubit quantum operations  
$ \CE_A(\rho) =  A \rho A^\dagger$,  there exists a quantum operation
$ \CE_D(\rho) =  D \rho D^\dagger$ with $[D,\rho] = 0$ and $D$ positive,
of the same output entropy and smaller or equal distortion.
\end{thm1}
$\mathbf{Proof }  $ \space \space
We work in the basis  $\{ \ket{0}, \ket{1} \}$ in which $\rho$ is diagonal, so
$ \rho =  p_0 \ket{0} \bra{0} +  p_1 \ket{1} \bra{1} $ with $p_0 + p_1 = 1$
and $p_0 \geq p_1 $.                                                
It is easy to see that any complex matrix $A$ 
can be expressed as a 
product $ A = U D \rho^{1/2} V \rho^{-1/2}$, where $U$ and $V$ are unitary 
and $D$ is diagonal positive and hence commutes with $\rho$. 
This follows from applying the polar decomposition of any complex matrix
$B$, namely $ B = U \Delta V $. Here  $U$ and $V$ are unitary, 
$\Delta$ is diagonal positive with non-ascending elements, and we choose
$B = A \rho^{1/2} $ and $D = \Delta  \rho^{-1/2} $. 
Such a decomposition ensures that $A \rho A^\dagger = U 
( D \rho D^\dagger ) U^\dagger$,
so that $tr(A \rho A^\dagger) = tr(D \rho D^\dagger)$ and
$S(\hat{\CE}_A) = S(\hat{\CE}_D)$. In addition, since both $\Delta = D
\rho^{1/2}$ and $\rho^{1/2}$ are diagonal positive with non-ascending
elements, Lemma 1 asserts that
$|tr(A \rho)| \leq |tr(D \rho)|$. 
Combining the above with the single qubit distortion
formula 
\beq 
d_e(\rho,\CE_A) = 1 -  \frac{ | tr(A \rho) | ^2 }{tr(A \rho A^\dagger) } ,
\label{eq:e3} 
\eeq
we see that the operation $\CE_D$ has the same output entropy
but a distortion that is less than or equal to that of $\CE_A$, 
thus proving the statement of the Theorem.
$~\star$

\vspace{2mm}

\centerline{ {\scalebox{.53}{\includegraphics{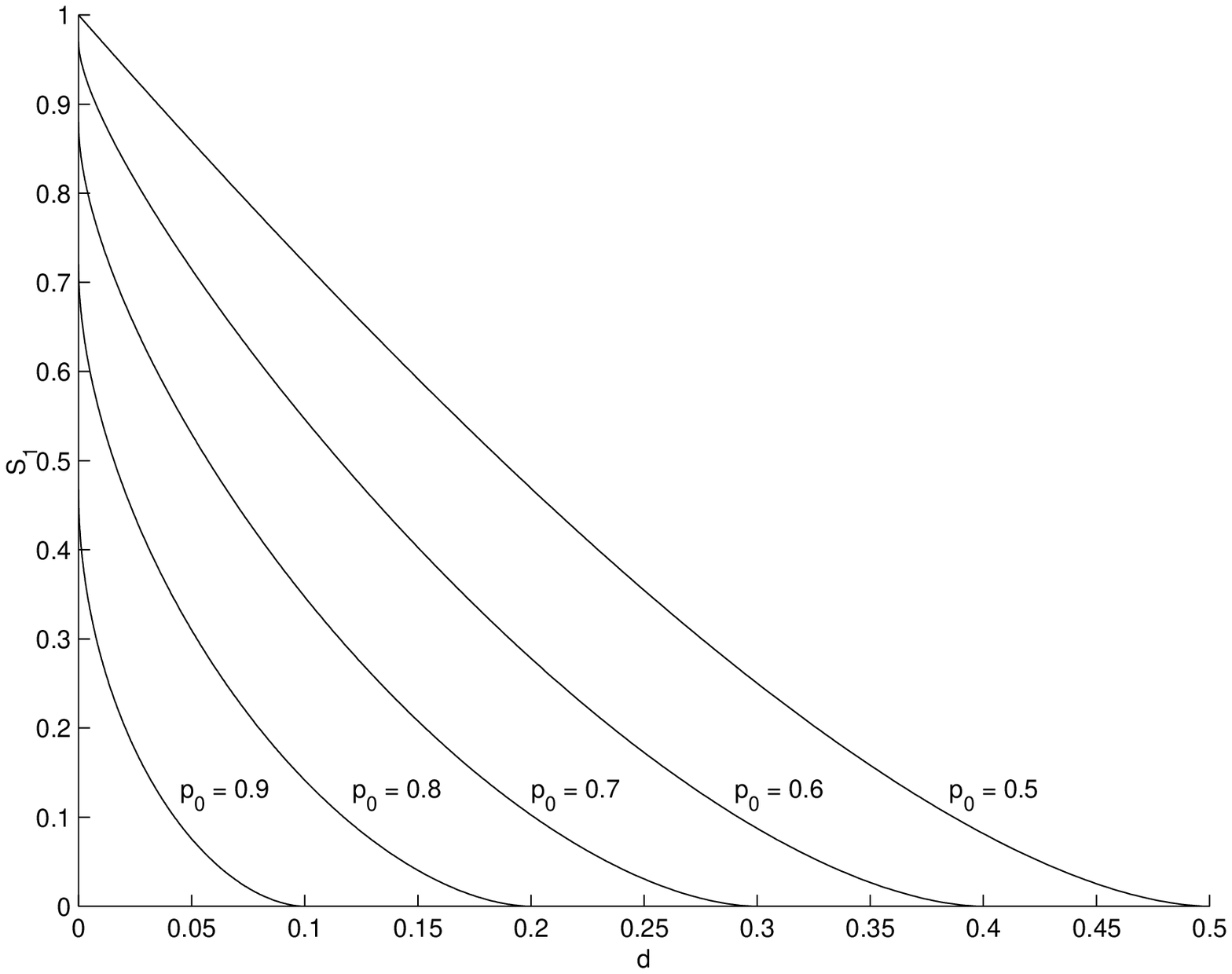}}}}
 
{\small {Fig. 1. Lower bound $S_1(d)$ on the single qubit
  rate-distortion function
for $p_0 = 0.5$, $0.6$, $0.7$, $0.8$ and $0.9$.}}

\vspace{4mm} 
\vskip 0.1in\noindent
Since $A$ is defined only up to a multiplicative constant,
Theorem 1 implies a complete parametrization for the unphysical $n = k = 1$ 
curve, which we denote here by $S_1(d)$. It is easy to see that
in the $\{ \ket{0}, \ket{1} \} $ basis the matrix
\beq
A =   \left(  \begin{array}{clcr}
             \cos \theta & 0  \\
                 0 &  \sin \theta
               \end{array}     \right)   ,~\theta \in [0, {\pi \over 4}],
\label{eq:e5}
\eeq
\vskip 0.1in\noindent
interpolates between the zero distortion limit $A = I$, where
$S = S(\rho)$, and the zero entropy limit $A = \ket{0} \bra{0} $,
where we replace the source with the pure ``best guess'' state  
$\ket{0} \bra{0}$.  

This curve, easily verified to be convex, is shown in Fig. 1 for several values 
of $p_0$. It is parametrized as

\beq
S_1(\Delta) =  h_2 \( \frac{p_0 (1 + \cos \Delta)}
{(p_0 + p_1) +  (p_0 - p_1) \cos \Delta} \) 
 , \,\,\,\,\,
d(\Delta) = \frac{p_0 p_1 (1 - \sin \Delta)}
{(p_0 + p_1) +  (p_0 - p_1) \cos \Delta}
\eeq

where $\Delta \in [0, {\pi \over 2}]$.
Here $h_2(\lambda) \equiv - \lambda \log_2(\lambda) - 
(1-\lambda) \log_2 (1 -\lambda) $ is the Shannon binary 
entropy function.
When $p_0 = {1 \over 2}$ the above simplifies to 
$S_1(d) = h_2(\frac{1}{2} + \sqrt{d(1-d)})$.
$S_1(d)$ serves as a lower bound for $R_1(d)$
since, for any decomposition
of unity $ \sum_i A_i^\dagger A_i = I$ and  
$\lambda_i = tr(\CE_{A_i}(\rho)) $, we have
\beq
\bar{S} = \sum_{i=1}^k \lambda_i  
\space S \( \hat{\CE}_{A_i}(\rho) \) 
\geq    \sum_{i=1}^k \lambda_i \space S_1(d_e(\rho ,
\CE_{A_i}))
\geq   S_1 \( \sum_{i=1}^k \lambda_i d_e(\rho,
\CE_{A_i}) \) 
\label{eq:k=1}
\eeq
by the convexity of $S_1(d)$. 
In the case of $p_0 = {1 \over 2}$, due to isotropy this lower 
bound is attainable with $k = 2$, 
\beq
A_1 =   \left(  \begin{array}{clcr}
                 \cos \theta  & 0  \\
                 0 &  \sin \theta 
               \end{array}     \right), \,
A_2 =   \left(  \begin{array}{clcr}
                 \sin \theta  & 0  \\
                 0 &  \cos \theta 
               \end{array}     \right),             
                 \, \theta \in [0, {\pi \over 4}]
\label{eq:yin1}
\eeq
\vskip 0.1in\noindent
The case $p_0 > {1 \over 2}$ is not as obvious. 
First we would like to show that $k = 2$ suffices.
We fix $A_1$ and vary $A_i$ , $2 \leq i \leq k$ . We
use Lagrange multipliers and seek the minimum of
\beq
\sum_{i=2}^k tr(A_i \rho A_i^\dagger) S\( {A_i \rho A_i^\dagger} \over
    {tr(A_i \rho A_i^\dagger)} \) 
    - \mu \sum_{i=2}^k  |tr(A_i \rho)|^2
    - \sum_{i=2}^k tr(\Lambda A_i^\dagger A_i)
\eeq
\vskip 0.1in\noindent
Differentiating $\bar{S}$ with respect to $A_i$ and $A_i^\dagger$ 
and setting this to zero, we obtain
an equation involving only $A_i$, $A_i^\dagger$, $\mu$ and $\Lambda$,
so evidently a solution is obtained for $A_2 = \dots = A_k$. This
has the same effect as retaining only $A_2$, so $k = 2$ includes
natural solutions to the extremum problem.  Motivated by the $p_0 = {1 \over 2}$ 
  case, we conjecture that the global \emph{minimum} is among them.
\vskip 0.1in\noindent
Restricting attention to $k = 2$, we concentrate on the 
case where $A_1$ and $A_2$ are diagonal and use the parametrization
\beq
     A_1 =   \left(  \begin{array}{clcr}
                 \cos \alpha  & 0  \\
                 0 &  \cos (\alpha + \Delta) 
               \end{array}     \right)~   , 
       A_2 =   \left(  \begin{array}{clcr}
                 \sin \alpha  & 0  \\
                 0 &  \sin (\alpha + \Delta)
               \end{array}     \right)             
                 , \Delta \in [0, {\pi \over 2}]
\label{eq:a1a2}
\eeq 
and $d = 2 p_0 p_1 (1 - \cos \Delta) $.
Here $\alpha$ is function of $\Delta$ such that
\beq 
\bar{S} = \sum_{i=1}^2 tr(A_i \rho A_i^\dagger) S\( {A_i \rho A_i^\dagger} \over
{tr(A_i \rho A_i^\dagger)} \) 
\label{eq:sbar}
\eeq
is maximized. Differentiating with respect to $\alpha$, we arrive at
\[ 2 p_0 p_1 \sin \Delta \left( 
\log_2 {\small \( \frac {p_1 \cos^2 (\alpha + \Delta)}{p_0 \cos^2 \alpha } \)}
 {\small \frac{\cos \alpha \cos (\alpha + \Delta)}
{p_0 \cos \alpha + p_1 \cos (\alpha + \Delta)}} \right. \]
\[ \left. + \log_2 
{\small \( \frac {p_1 \sin^2 (\alpha + \Delta)}
{p_0 \sin^2 \alpha } \)}
{\small \frac{\sin \alpha \sin (\alpha + \Delta)}
{p_0 \sin \alpha + p_1 \sin (\alpha + \Delta)} } \right) \]
\[ + ( p_0 \sin 2 \alpha + p_1 \sin 2(\alpha + \Delta) )
  \( h_2\( \frac{p_0 \sin^2 \alpha}{p_0 \sin^2 \alpha + p_1 \sin^2 (\alpha + \Delta)} \)
  - h_2 \( \frac{p_0 \cos^2 \alpha}{p_0 \cos^2 \alpha + p_1 \cos^2 (\alpha + \Delta)} \) \)
   = 0, \]
\label{eq:goo}
\vskip 0.1in\noindent
which we solve numerically.
The function $\alpha(\Delta)$ is plotted
in Fig. 2 for several values of $p_0$. 
We also plot the corresponding rate-distortion curves
in Fig. 3. The curves are convex and approach $d_{max} = 2 p_0 p_1$ with
zero slope.
Note that the $p_0 = {1 \over 2}$ solution is precisely the one obtained
previously, namely $S_1(d)$.
\vspace{2mm}

\centerline{ {\scalebox{.53}{\includegraphics{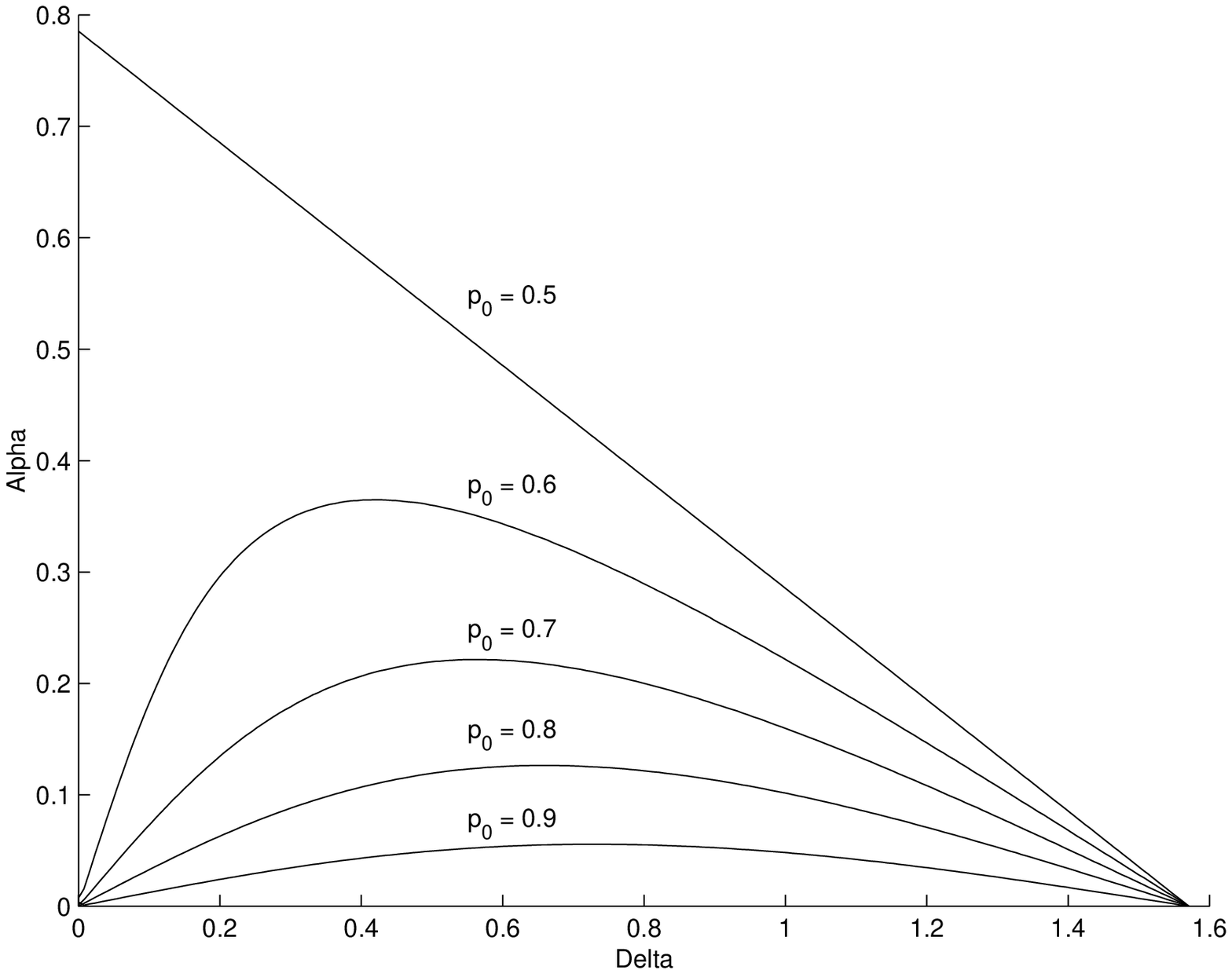}}}}
\begin{center} 
{\small {Fig. 2. The function $\alpha(\Delta)$ that solves (\ref{eq:goo})
plotted for $p_0 = 0.5$, $0.6$, $0.7$, $0.8$ and $0.9$.}}
\end{center}
\vspace{2mm}
\vskip 0.1in\noindent
Now we show that this diagonal solution is optimal with respect to local
perturbations of the $\{A_i\}$.   
Recall that we wish to find the optimal tradeoff between $\bar{S}$
and  $ d = 1 - \sum_i |tr(A_i \rho)|^2 $ under the constraint 
$ \sum_i A_i^\dagger A_i = I$.
Notice that both $\bar{S}$ and the trace preserving condition are invariant
under the transformation $A_i \rightarrow U_i A_i$ where $U_i$ are unitary
matrices. Furthermore $|tr(U_i A_i \rho)| \leq |tr(A_i \rho)|$ 
when $A_i \rho$ is positive (see Lemma 2 below), and we may always pick $U_i$ to
achieve this upper bound. This can be seen from the polar decomposition 
$A_i \rho = V_i D_i W_i$  and choosing  $U_i = (V_i W_i)^{-1}$.
Therefore we restrict attention to positive $A_i \rho$ and use
a new parametrization:
\beq
A_1 =  f \left(  \begin{array}{clcr}
        \frac{ \lambda \cos \theta}{p_0}  &  {x \sin \theta  \over p_1}  \\
    { {x^* \sin \theta } \over p_0}   &   \frac{(1 - \lambda) \cos \theta }{p_1}
               \end{array}     \right) , 
  A_2 =  f \left(  \begin{array}{clcr}
        \frac{ \mu \sin \theta }{p_0}  &  -{x \cos \theta  \over p_1}  \\
      -{ {x^* \cos \theta } \over p_0}   &   \frac{(1 - \mu) \sin \theta }{p_1}
               \end{array}     \right)                
       \label{eq:yin2}
\eeq
\vspace{2mm}
in terms of $\theta$ and complex $x$. Here $\lambda$ and
$\mu$ are functions of $|x|$ determined by the conditions
\beq
\lambda^2 \cos^2 \theta + \mu^2 \sin^2 \theta = {p_0^2 \over f^2} - |x|^2
\eeq 
\beq
(1 - \lambda)^2 \cos^2 \theta + (1 -\mu)^2 \sin^2 \theta = 
{p_1^2 \over f^2} - |x|^2
\eeq 
\vskip 0.1in\noindent
and $d = 1 - f^2$.
We see from the expansion about $x = 0$ that $\lambda$ and $\mu$ 
are both quadratic
in $|x|$. It is also easy to see that the traces and determinants of
the $A_i \rho A_i^\dagger$ (and hence the eigenvalues) 
also have no terms linear in $x$.
Expanding to second order about the optimal diagonal solution, we verify
that $\bar{S}$ is indeed at a local minimum with respect to varying $x$.
We thus conclude our argument that the $n = 1$ rate-distortion curves
$R_1(d)$ are those depicted in Fig. 3.
\vspace{2mm}
\centerline{ {\scalebox{.53}{\includegraphics{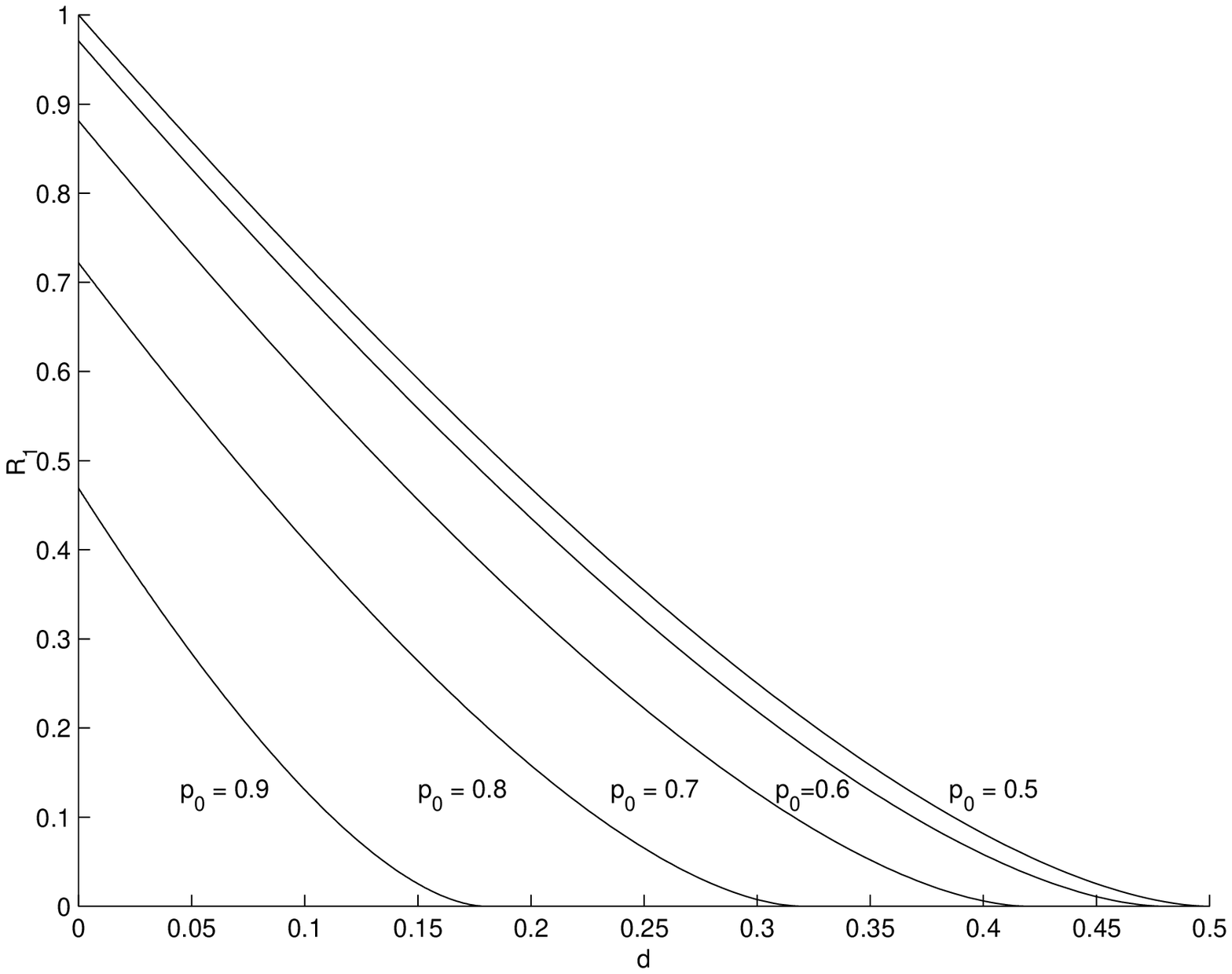}}}}
\begin{center} 
{\small {Fig. 3. The single qubit rate-distortion function $R_1(d)$
plotted for $p_0 = 0.5$, $0.6$, $0.7$, $0.8$ and $0.9$.}}
\end{center}
\vspace{4mm} 
\section{The rate-distortion function for general $n$ }
Now we move to the general $n$ case and argue that we 
cannot do any better than $R_1(d)$. 
We have $n$ qubits with joint density operator $\rho^{\otimes n}$,
and we consider appropriate combinations of quantum operations 
$ \CE_A(\rho^{\otimes n}) =  A ( \rho^{\otimes n} ) A^\dagger$.
We work in the basis $\CB^n = \{ \ket{0} , \ket{1} \}^n $ 
with  $\ket{0}$ and $\ket{1}$ defined as before.
In this basis the system operator $A$ is given by
\beq
  A =  \left(  \begin{array}{clcr}
                 B & K   \\
                 L & C
               \end{array}     \right)~, 
\eeq
\vskip 0.1in\noindent
where the $B$, $K$, $L$ and $C$ are $2^{n-1} \times 2^{n-1}$ matrices acting 
on the last $n-1$  qubits.
It is easy to verify that the restriction  $\CE^>$ of  $\CE$ to
the last $n-1$ qubits is given by the set 
$ \{ \sqrt p_0 B ,\sqrt p_1 K ,\sqrt p_0  L,
\sqrt p_1  C \} $ of operation elements.
 
We first restrict attention to processes with $A$ diagonal in the 
$\CB^n$ basis.

\begin{thm2}
General $n$-qubit trace-preserving  
processes with operation elements \{$A_i$\} diagonal
in the $ \CB^n$ basis cannot perform below the single qubit rate-distortion 
curve $R_1(d)$. 
\end{thm2}
\noindent
$\mathbf{Proof }  $ \space \space 
We prove the theorem using induction on $n$. It is true for
$n=1$ by the results of the previous section. 
Let us now assume it holds for $n$, and then show its validity for $n+1$.
We work in the $\CB^{n+1}$ basis where $A_i$ is represented
by a $2^{n+1} \times 2^{n+1}$ dimensional matrix
\beq   
   A_i =   \left(  \begin{array}{clcr}
                  {1 \over  \sqrt{p_0}} B_i  &  \\
                    &  {1 \over  \sqrt{p_1}} C_i  \\
               \end{array}     \right)
\label{eq:e11}
\eeq   
\vskip 0.1in\noindent
with $B_i$ and $C_i$ both diagonal $2^{n} \times 2^{n}$ matrices acting
on the last $n$ qubits. Then the projection of $\CE_{A_i}$ onto the last
$n$ qubits is 
$\CE^{>}_{A_i}(\rho^{\otimes n}) = B_i \rho^{\otimes n} B_i^\dagger
+ C_i \rho^{\otimes n} C_i^\dagger$.  We also have from (\ref{eq:e11}) that
\beq
   \CE_{A_i}(\rho^{\otimes n+1}) = \left(  \begin{array}{clcr}
                B_i \rho^{\otimes n} B_i^{ \dagger }  &    \\
                      &   C_i \rho^{\otimes n} C_i^{ \dagger }
               \end{array}     \right) 
\eeq
\vskip 0.1in\noindent
Then the normalized projection of $\CE_{A_i}$ onto the first qubit is
\beq   
  \hat{\CE}^{1}_{A_i}(\rho) =  \left(  \begin{array}{clcr}
                       \lambda_i  &    \\
                      &    1 - \lambda_i
               \end{array}     \right)  
\label{eq:e12}
\eeq
where $\lambda_i = {tr(\CE_{B_i}(\rho^{\otimes n}))}/
{tr(\CE_{A_i}(\rho ^{\otimes n+1}))}$.

The average distortion  associated with the coding procedure defined by
the $\{A_i \}$  is   

\beq
  d = {n \over {n+1}} d^> + {1 \over {n+1}} d^1
\label{eq:de}
\eeq

where

\beq
   d^> =  \sum_i tr(\CE_{B_i}(\rho ^{\otimes n}))
 d_e(\rho ^{\otimes n},\CE_{B_i}) +
 tr(\CE_{C_i}(\rho ^{\otimes n}))
 d_e(\rho ^{\otimes n},\CE_{C_i}) 
\eeq

and
 
\beq
      d^1 = \sum_i d_e(\rho,\CE^{1}_{A_i})
\eeq

Using the simple identity

\beq
   S( \lambda \rho_1 \oplus (1 -\lambda) \rho_2 ) = 
  \lambda S(\rho_1) +  (1 -\lambda) S(\rho_2) + h_2(\lambda)  
\eeq

we find that

\beq
    S (\hat{\CE}_{A_i}(\rho ^ {\otimes n+1 } ) = 
\lambda_i S ( \ \hat{\CE}_{B_i}(\rho^ {\otimes n } ) ) +
(1 - \lambda_i) S ( \hat{\CE}_{C_i}(\rho^ {\otimes n }) ) +  h_2(\lambda_i).
\label{eq:e13}
\eeq

Hence:
 
\begin{eqnarray} 
\lefteqn{ {1 \over {n+1}} \sum_i tr(\CE_{A_i}(\rho ^{\otimes n+1}))
 S (\hat{\CE}_{A_i}(\rho ^ {\otimes n+1 } ))} \nonumber \\
 & = & {n \over {n+1}} \( 
 {1 \over n}  \sum_i tr(\CE_{B_i}(\rho ^{\otimes n}))
 S ( \ \hat{\CE}_{B_i}(\rho^ {\otimes n } ) ) +
 tr(\CE_{C_i}(\rho ^{\otimes n}))
 S ( \ \hat{\CE}_{C_i}(\rho^ {\otimes n } ) )  \) \nonumber\\ 
 & + &  {1 \over {n+1}}  \sum_i tr( {\CE}^{1}_{A_i}(\rho))
   S( \hat{\CE}^{1}_{A_i}(\rho)) \nonumber \\
 &  \geq & {n \over {n+1}} R_1(d^>) +
 {1 \over {n+1}} R_1(d^1) 
  \, \geq \,  R_1(d)
\end{eqnarray}
\vskip 0.1in\noindent
The equality comes from (\ref{eq:e12}),(\ref{eq:e13})  
and the fact that $tr(\CE_{A_i}(\rho ^{\otimes n+1}))  = 
tr( \hat{\CE}^{1}_{A_i}(\rho))$, the first 
inequality comes from the inductive hypothesis, and the second inequality
is a consequence of convexity of $R_1(d)$ and (\ref{eq:de}).
Hence, the rate for $\{A_i\}$ is greater than or equal to 
$R_1(d)$ at the same distortion, as claimed. $\star$
\vspace{4mm} 

Finally, it remains to show that for general $n$ diagonal processes are optimal.
This may be shown exactly in the case $p_0 = {1 \over 2}$
due to its many simplifying features.
We begin with two lemmas.
 
\begin{lemma'}
Given matrices $\{Y_i \}$   with 
$ \sum_i Y_i^\dagger Y_i  =  I$ 
and positive $D$, the inequality   
$\sum_i |tr(Y_i D)|^2  \leq |tr(D)|^2$ holds.
\end{lemma'}
\noindent
$\mathbf{Proof }  $ \space \space 
We use the fact that $D = \sqrt{D D^\dagger}$ for $D$ positive 
and employ the Cauchy-Schwartz inequality (\ref{eq:CS}) to write
\beq
\sum_i |tr(Y_i D)|^2 = \sum_i |tr( ( Y_i \sqrt{D}) \sqrt{D^\dagger })|^2 
\leq \sum_i tr( Y_i D Y_i^\dagger) tr(D) =  |tr(D)|^2
\eeq
\noindent
The last equality comes from the cyclicity and linearity of trace. $\star$
\begin{lemma'2}
Given operators  $\{Y_i \}$  acting on n qubits
with $ \sum_i Y_i^\dagger Y_i =  I$  
and  positive $D$, diagonal in the $\CB^n$ basis, we have the inequality
\[ 
\sum_i tr(\CE_{Y_i D}(\rho ^{\otimes n})) d_e(\rho ^{\otimes n},\CE_{Y_i D})
\geq tr(\CE_{D}(\rho ^{\otimes n}))    d_e(\rho ^{\otimes n},\CE_{D}). \]
\end{lemma'2}
\vskip 0.1in\noindent
$\mathbf{Proof }  $ \space \space
We again use induction. The $n=1$ case follows from
Lemma 2. Assuming the Lemma holds for $n$ we prove it for $n+1$.
Consider $2^{n+1} \times 2^{n+1}$ dimensional matrices $\{Y_i \}$,
and let
\beq
     Y_i =  \left(  \begin{array}{clcr}
                 E_i & F_i   \\
                 G_i & H_i
               \end{array}     \right) \,\,\,\,\,\,\,\,\,\,\,
  D =   \left(  \begin{array}{clcr}
                 {1 \over  \sqrt{p_0}} D_0 &    \\
                  &  {1 \over  \sqrt{p_1}} D_1
               \end{array}     \right)
\eeq
\vskip 0.1in\noindent
with $E_i$ etc. of dimension $2^{n} \times 2^{n}$. 
$ \sum_i Y_i^\dagger Y_i =  I$ implies that
\beq
   \sum_i \( E_i^\dagger E_i + G_i^\dagger G_i \) = I
\label{eq:e14} 
\eeq
\vskip 0.1in\noindent
and similarly for $F_i$ and $H_i$.
The restriction $\CE^>_{Y_i D}$of 
$\CE_{Y_i D}$ onto the last
$n$ qubits is described by the set 
$\{ E_i D_0 , F_i D_1 , G_i D_0 ,H_i D_1 \}$. Then
\begin{eqnarray} 
\sum_i tr(\CE^>_{Y_i D}(\rho ^{\otimes n})) 
d_e(\rho ^{\otimes n},\CE^>_{Y_i D}) & = & 
{\sum_i  
tr(\CE_{E_i D_0}(\rho ^{\otimes n})) d_e(\rho ^{\otimes n},\CE_{E_i D_0})
+ tr(\CE_{F_i D_1}(\rho ^{\otimes n})) d_e(\rho ^{\otimes n},\CE_{F_i D_1})}
\nonumber \\
& + & tr(\CE_{G_i D_0}(\rho ^{\otimes n})) d_e(\rho ^{\otimes n},\CE_{G_i D_0})
+ tr(\CE_{H_i D_1}(\rho ^{\otimes n})) d_e(\rho ^{\otimes n},\CE_{H_i D_1}) 
\nonumber \\
& \geq &  \sum_i  
tr(\CE_{D_0}(\rho ^{\otimes n})) d_e(\rho ^{\otimes n},\CE_{D_0})
+ tr(\CE_{D_1}(\rho ^{\otimes n})) d_e(\rho ^{\otimes n},\CE_{D_1})  
\nonumber \\
& = & tr(\CE^>_{D}(\rho ^{\otimes n})) 
d_e(\rho ^{\otimes n},\CE^>_{D})
\end{eqnarray} 
\vskip 0.1in\noindent
The inequality comes from the inductive hypothesis and (\ref{eq:e14}).
Finally, this result is invariant under permutations of the qubits;
averaging over all permutations yields 
\beq
    \sum_i tr(\CE_{Y_i D}(\rho ^{\otimes n+1})) 
 d_e(\rho ^{\otimes n+1},\CE_{Y_i D}) \geq
 tr(\CE_{D}(\rho ^{\otimes n+1})) 
 d_e(\rho ^{\otimes n+1},\CE_{D})
\eeq
\vskip 0.1in\noindent
This proves the Lemma.    $\star$
\begin{thm3}
General n-qubit processes cannot perform below the single qubit 
entropy-distortion curve $S_1(d)$ in the case of isotropic sources 
$(p_0 = {1 \over 2})$. 
\end{thm3}
\vskip 0.1in\noindent
$\mathbf{Proof }  $ \space \space 
This is an immediate consequence of Lemma 3.
We ignore the trace preserving condition for
the time being and consider
$ \CE_A(\rho^{ \otimes n} ) =  A (\rho^{ \otimes n}) A^\dagger$.
Then we use the
polar decomposition  $ A = U D V$ with $U$ and $V$ unitary
and $D$ diagonal positive. Using the fact that 
$\rho = {1 \over 2} I$, it is easy to see that
$tr(\CE_A(\rho^{ \otimes n})) = tr(\CE_D(\rho^{ \otimes n} ))$ ,
$S(\hat{\CE}_A (\rho^{ \otimes n})) = 
S(\hat{\CE}_D (\rho^{ \otimes n}  ))$
and $d_e(\rho ^{\otimes n},\CE_{A}) = d_e(\rho ^{\otimes n},\CE_{V U D})$.
Then from Lemma 3 with $m = 1$ and $Y_1 = V U$, we get 
$d_e(\rho ^{\otimes n},\CE_{A})
\geq d_e(\rho ^{\otimes n},\CE_{D}) $. Therefore, there is a diagonal map that 
does at least as well as $\CE_A$.
From a trivial variation on Theorem 2 
(note that the trace-preserving condition plays no role in the proof),
this diagonal map cannot do better than the $n = k = 1$ curve $S_1(d)$ which
is attainable for $p_0 =  {1 \over 2}$. 
Having established that the optimal $\CE_A$ yields the convex $S_1(d)$,
using the same argument as in (\ref{eq:k=1}) we see that reintroducing
the trace-preserving condition does not affect our result. Hence the
Theorem is proved. $\star$
\vskip 0.1in\noindent
We conjecture that the theorem also holds for the case $p_0 > {1 \over 2}$,
and we now present some evidence to support this conjecture. It again 
suffices to show that 
diagonal processes are optimal for general $n$.

$\bullet$ Consider perturbing a process
defined by $2^{n} \times 2^{n}$ dimensional diagonal matrices $\{A_i\}$ with 
$\sum_i A_i^\dagger A_i = I$  by a general 
matrices $\{Q_i\}$ with diagonal elements all
equal to zero. It is easy to see that to \emph{linear} order 
the trace-preserving condition still holds, 
and both average entropy and distortion remain unchanged.
Hence, all diagonal processes are local extrema with respect to off-diagonal
perturbations. 

$\bullet$ In Theorem 2 we never used the fact that 
$B_i$ and $C_i$ were diagonal, so a more general class of operators given by
(\ref{eq:e11}), in $\CB^n$ or any other basis obtained by permutations of
the qubits, lies above the $R_1(d)$ curve.  
 
$\bullet$ A straightforward modification of Theorem 3 
shows that diagonal processes 
${D_i}$ do better than ${U_i D_i}$, 
where $U_i$ is any unitary operator (note that the trace preserving 
condition still holds).

$\bullet$ By iterating the argument preceding Theorem 2, the restriction of
a general $n$-qubit operation onto a single qubit involves 
$2^{n-1}$ operation elements which greatly increases the entropy exchange
with the environment of that qubit. 
Essentially, individual qubits act as the environment for each other,
and entangling them creates noise.
On the other hand, as in classical information theory, the
benefit of entangling (correlating) the qubits is a reduction in entropy 
since  $ S(\CE(\rho^{\otimes n})) \leq \sum_\alpha S(\CE^{\alpha}(\rho)) $
where $\CE^{\alpha}$ is the restriction of $\CE$ to the $\alpha$th qubit.
There is a competition between these two effects, and the former wins, 
as we have proven rigorously for $p_0 = {1 \over 2}$. 
In this sense, however, there is 
nothing special about $p_0 = {1 \over 2}$. If anything, we would expect
the entropy to be even harder to reduce via quantum operations 
for $p_0 > {1 \over 2}$
than for $p_0 = {1 \over 2}$ because it is lower to start with. 
\section{Physical realization of the $R(d)$ curve}
We now elaborate on how our coding procedure may be realized physically.
For the lossy part of the coding we need to provide an 
ancilla qubit in a definite
state. We first apply a unitary transformation entangling the ancilla with the
source qubit, and then measure the ancilla. 
In the basis $\{ \ket{0}_A \ket{0}_Q, 
\ket{0}_A\ket{1}_Q,  \ket{1}_A\ket{0}_Q , \ket{1}_A \ket{1}_Q \}$, 
the unitary transformation is given by the matrix
\beq
 U =  \left(  \begin{array}{clcr}
                \cos \alpha  &   & -\sin \alpha  &   \\
                      & \cos(\alpha + \Delta) &  & -\sin(\alpha + \Delta)   \\
                   \sin \alpha & & \cos \alpha &  \\
                   & \sin (\alpha + \Delta) & &  \cos(\alpha + \Delta)   \\
               \end{array}     \right)
\eeq
\vskip 0.1in\noindent
with $ \Delta \in [0, {\pi \over 2} ]$ and $\alpha = \alpha(\Delta)$
as defined before. The ancilla is prepared in the $\ket{0}_A$
state so that the initial density operator for the ancilla-source system is
\beq
\Xi  =  \left(  \begin{array}{clcr}
                 \rho & \mathbf{0}  \\
                  \mathbf{0}   &  \mathbf{0}
               \end{array}     \right) 
\eeq
\vskip 0.1in\noindent
Then 
\beq
  U \Xi U^\dagger  = \left(  \begin{array}{clcr}
                  A_1 \rho A_1^\dagger &  A_1 \rho A_2^\dagger  \\
                    A_2 \rho A_1^\dagger &  A_2 \rho A_2^\dagger 
                \end{array}  \right) ~,
\eeq
where $A_1$ and $A_2$ are as defined in (\ref{eq:a1a2}).
We then measure the ancilla qubit. If the outcome is $\ket{0}_A$,
we know the map $\rho \rightarrow\hat{\CE}_{A_1}(\rho)$ 
has been performed
and we label the qubit as belonging to type 1. Similarly, if the outcome is
$\ket{1}_A$, we know the map 
$\rho \rightarrow\hat{\CE}_{A_2}(\rho)$ has transpired and label the qubit
to be of type 2. In the end
we perform two Schumacher encodings, one on all the bits of the first type 
and a separate one on all the bits of the second type.
When decoding, we need information about the sequence of
operations performed. The rate of classical information required
for this is $r = h_2(tr(A_1 \rho A_1^\dagger))$.
These classical rates are plotted for several values of $p_0$ in Fig. 4.
\section{Discussion}
We have shown that for the distortion measure in question
and when allowed an unrestricted classical side channel, optimum quantum
rate-distortion codes are separable 
into a lossy part involving single qubit operations followed by
standard Schumacher lossless coding of large blocks of qubits.

Our result has the following interpretation: the rate-distortion
curve is achieved by quantum operations that 
produce no entropy exchange with the environment of any individual qubit.
We do not expect zero entropy exchange to be optimal
for more general distortion measures.  Since our distortion
measure, which is based on the concept of entanglement fidelity, 
emphasizes preserving the state of $RQ$, it
forbids any increase of the entropy of $RQ$ which means it forbids
any entropy exchange.
We also do not believe $n = 1$ to be optimal when restrictions on $r$ are 
imposed since, as remarked in Section 2, the entropy exchange
is positive as long as there is uncertainty in the value of the 
index random variable $I$.

\vspace{4mm}

\centerline{ {\scalebox{.53}{\includegraphics{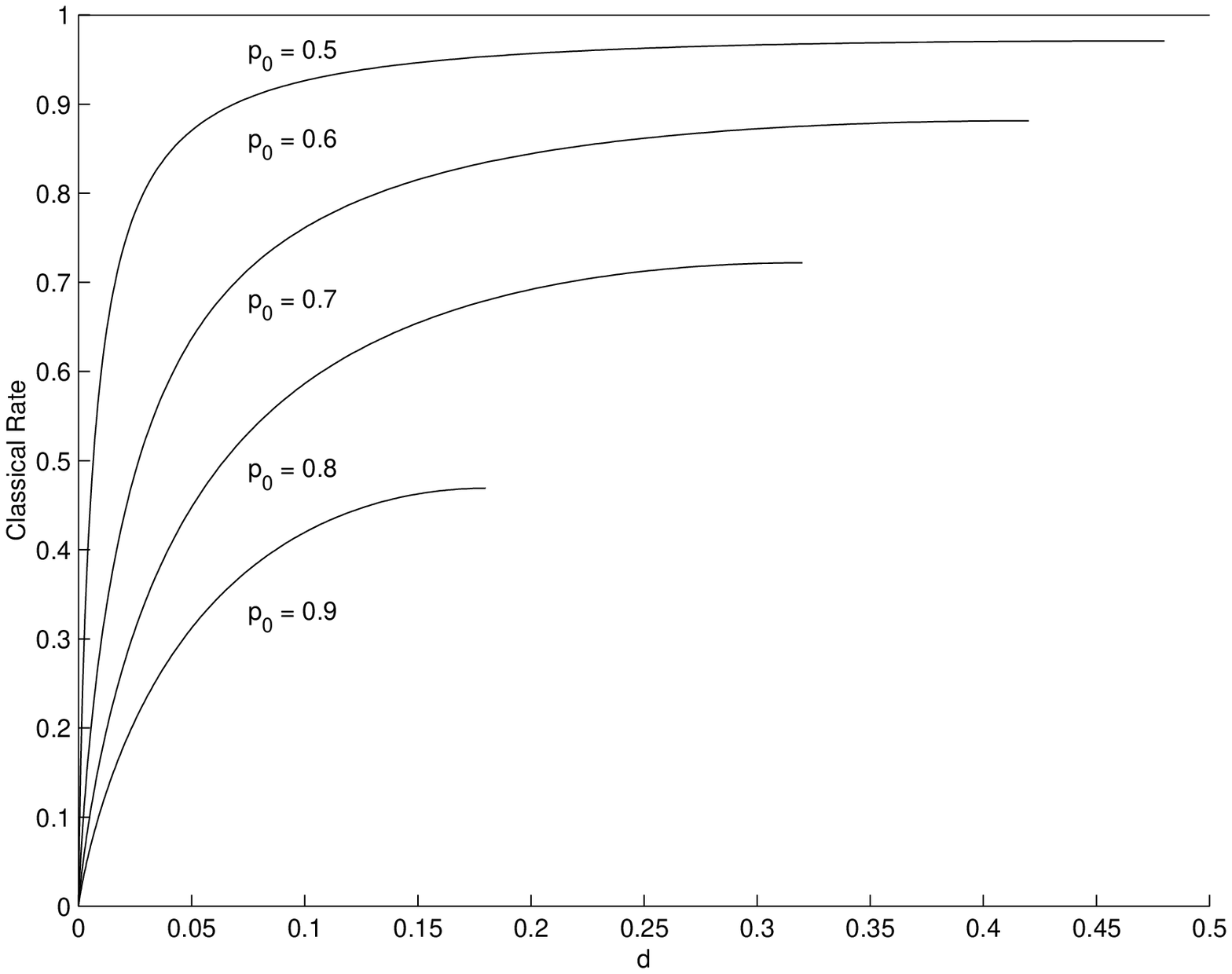}}}}
{\small {Fig. 4. The classical information rate needed to decode 
vs. $d$ for $p_0 = 0.5$, $0.6$, $0.7$, $0.8$ and $0.9$.}}
\vskip 0.1in
Let us examine the action of our quantum map on
normalized pure states.
If we picture $\ket{0}$ and $\ket{1}$ as orthogonal vectors then,
depending on which of the two operations has been performed, the map 
rotates our pure state vector toward $\ket{0}$ or toward $\ket{1}$.
The source is biased toward $\ket{0}$, which it produces with a
higher probability than $\ket{1}$. The first type of operation
produces qubits biased
even more toward $\ket{0}$, hence causing a decrease in entropy.
The second type does the opposite and perhaps even increases the entropy
for $p_0 > {1 \over 2}$; however, it has to 
occur a  certain fraction of the time in order to obey the trace-preserving 
condition, which says that the total probability of performing
\emph{some} operation must be equal to $1$ regardless of the
input state.  On average, the entropy does decrease, while
the discrepancy between
the initial and final state increases. The $R(d)$ curve is thus swept out.  
 
Notice that our quantum $R(d)$ curve first falls to $R = 0$
at $d_{max} = 2 p_0 p_1$, as opposed to the classical value $d_{max} = p_1$
associated with reconstructing the source bit 
with the best guess at its value.
This, too, is due to our choice of fidelity measure: replacing the original
qubit with a fresh one prepared in the state $\ket{0}$ destroys the
entanglement with the original reference system. The best we can do 
is project onto $\ket{0}$ with probability $p_0$ and otherwise
project onto $\ket{1}$.
   
We do not expect a general expression
resembling the classical prescription (\ref{eq:e0}) 
for the rate-distortion function that is 
valid for all distortion measures to exist for quantum
rate-distortion. Our reason
for this lies in the richness of distortion measures which vary in their
degree of "quantumness". The one we have used based on entanglement
fidelity evidently has a highly quantum nature. On the other hand,
we could view $\rho$ as being realized by a specific ensemble like 
$\CQ = \{ (\ket{0}, p_0) ,  (\ket{1}, p_1) \}$, 
and use as our distortion
measure the corresponding average pure state distortion measure 
$\overline d(\CQ^{n} , \CD^{(n)} \circ \CC^{(n)} ) $ based on the
average pure state fidelity $ {\overline F}(\CQ, \CE) $, namely
\beq
  {\overline F}(\CQ, \CE) = p_0 \bra{0} \CE(\ket{0}\bra{0}) \ket{0}
+ p_1 \bra{1} \CE(\ket{1}\bra{1}) \ket{1}
\eeq
\vskip 0.1in\noindent
Here we are able to attain zero distortion merely by sending classical 
information -- the measurement results in the $\{\ket{0},\ket{1}\}$ 
basis.  If we do not allow storing classical information, then the 
appropriate cross section of the rate-distortion manifold 
becomes the classical rate-distortion function for the Hamming 
measure, namely $ R(d,0) = S(\rho) - h_2(d) $. 

One could also investigate more general ensembles, as well as  
distortion measures tied to 
specific quantum cryptography protocols. Finally, the work presented here
naturally generalizes to systems with more than two degrees of freedom.
\paragraph{Acknowledgement} 
 
We thank Konrad Banaszek, Howard Barnum, David Mermin, Ian Walmsley 
and anonymous referees for valuable comments
and particularly for pointing out problem formulation inadequacies in earlier
versions of the manuscript. This research was 
supported in part by
the DoD Multidisciplinary University Research
Initiative (MURI) program administered by the Army Research Office under
Grant DAAD19-99-1-0215 and NSF Grant CCR-9980616.


\begin{thebibliography}{99}

\bibitem{berger} T. Berger, {\it Rate Distortion Theory}, Prentice Hall 
(1971)

\bibitem{nono} B. Schumacher, ``Quantum coding'',
{\it Phys.Rev.A }{\bf 51},
2738 (1995); R. Jozsa and B. Schumacher, 
``A new proof of the quantum noiseless coding theorem'',
{\it J. Mod. Optics}{\bf 41}, 
2343 (1994)  
 
\bibitem{noisy} S. Lloyd, 
``Capacity of the noisy quantum channel'', {\it Phys. Rev. A }{\bf 55},
1613 (1996) 

\bibitem{bj} H.Barnum, P.Hayden, R.Jozsa and A.Winter,
``On the reversible extraction of classical information 
from a quantum source'', LANL preprint quant-ph/0011072

\bibitem{sn} B. Schumacher and M. A. Nielsen,
``Quantum data processing and error correction'',
{\it Phys. Rev. A }{\bf 54},
2629 (1996)

\bibitem{barnum} H. Barnum, ``Quantum rate-distortion coding'',
{\it Phys. Rev. A} {\bf 62}, 42309 (2000)

\bibitem{sch} B. Schumacher, 
``Sending entanglement through noisy quantum channels'',
{\it Phys. Rev. A }{\bf 54},
 2614  (1995)
  
\bibitem{bns} H. Barnum, M. A. Nielsen and B. Schumacher, 
``Information transmission through a noisy quantum channel'',
{\it Phys. Rev. A } 
{\bf 57}, 4153 (1998)
  
\bibitem{chuang} I. L. Chuang and D. S. Modha,
``Reversible arithmetic coding for quantum data compression'',
{\it IEEE Trans. IT } {\bf 46},
1104 (2000)

\bibitem{wehrl} A. Wehrl,
``General properties of entropy'',
{\it Rev.Mod.Phys } {\bf 50},
221 (1978)

\end{thebibliography}
\end{document}